\documentclass[conference]{IEEEtran}
 \usepackage{amsmath,amssymb}
 \usepackage{subfigure}
 \usepackage{graphicx,graphics,color,psfrag}
 \usepackage{cite,balance}
 \usepackage{algorithm}
 \usepackage{accents}
 \usepackage{amsthm}
 \usepackage{bm}
 \usepackage{url}
 \usepackage{algorithmic}
 \usepackage[english]{babel}
 \usepackage{multirow}
 \usepackage{enumerate}
 \usepackage{cases}
 \usepackage{stfloats}
 \usepackage{dsfont}
 \usepackage{color,soul}
 \usepackage{amsfonts}
  \usepackage{tcolorbox}

 \usepackage{cite,graphicx,amsmath,amssymb}
 \usepackage{subfigure}
 \usepackage{fancyhdr}
 \usepackage{hhline}
 \usepackage{graphicx,graphics}
 \usepackage{array,color}
 \usepackage{amsmath}
 \usepackage{amsthm}

\newtheorem{remark}{Remark}
\newtheorem{theorem}{Theorem}

\newtheorem{assumption}{Assumption}

\include{header}

\IEEEoverridecommandlockouts

\begin{document}

\title{Optimized Power Control for Over-the-Air Federated Edge Learning}

\author{\IEEEauthorblockN{Xiaowen~Cao\IEEEauthorrefmark{1}\IEEEauthorrefmark{3},
Guangxu~Zhu\IEEEauthorrefmark{2},
Jie~Xu\IEEEauthorrefmark{3}, and
Shuguang~Cui\IEEEauthorrefmark{3}\IEEEauthorrefmark{2}
}
\IEEEauthorblockA{\IEEEauthorrefmark{1}School of Information Engineering, Guangdong University of Technology, Guangzhou, China}
\IEEEauthorblockA{\IEEEauthorrefmark{2}Shenzhen Research Institute of Big Data, Shenzhen, China}
\IEEEauthorblockA{\IEEEauthorrefmark{3}FNii and SSE, The Chinese University of Hong Kong (Shenzhen), Shenzhen, China \\
Email: caoxwen@outlook.com,~gxzhu@sribd.cn,~xujie@cuhk.edu.cn,~shuguangcui@cuhk.edu.cn}
}

\markboth{}{}
\maketitle

\setlength\abovedisplayskip{2pt}
\setlength\belowdisplayskip{2pt}

\begin{abstract}
\emph{Over-the-air federated edge learning} (Air-FEEL) is a communication-efficient solution for privacy-preserving distributed learning over wireless networks. Air-FEEL allows ``one-shot" over-the-air aggregation of gradient/model-updates by exploiting the waveform superposition property of wireless channels, and thus promises an extremely low aggregation latency that is independent of the network size. However, such communication efficiency may come at a cost of learning performance degradation due to the aggregation error caused by the non-uniform channel fading over devices and noise perturbation. Prior work adopted channel inversion power control (or its variants) to reduce the aggregation error by aligning the channel gains, which, however, could be highly suboptimal in deep fading scenarios due to the noise amplification. To overcome this issue, we investigate the power control optimization for enhancing the learning performance of Air-FEEL. Towards this end, we first analyze the convergence behavior of the Air-FEEL by deriving the optimality gap of the loss-function under any given power control policy. Then we optimize the power control to minimize the optimality gap for accelerating convergence, subject to a set of average and maximum power constraints at edge devices. The problem is generally non-convex and challenging to solve due to the coupling of power control variables over different devices and iterations. To tackle this challenge, we develop an efficient algorithm by jointly exploiting the \emph{successive convex approximation} (SCA) and trust region methods. Numerical results show that the optimized power control policy achieves significantly faster convergence than the benchmark policies such as channel inversion and uniform power transmission.
\end{abstract}

\section{Introduction}\label{sec:intro}
In the pursuit of ubiquitous intelligence envisioned in the future 6G networks, recent years have witnessed the spreading of {\it artificial intelligence} (AI) algorithms from the cloud to the network edge, resulting in an active area called {\it edge intelligence} \cite{Survey_FEEl,Debbah19}.
The core research issue therein is to allow low-latency and privacy-aware access to rich mobile data for intelligence distillation. To this end, a popular framework called {\it federated edge learning} (FEEL) is proposed recently, which distributes the task of model training over edge devices so as to reduce the communication overhead and keep the data-use locally \cite{Konecny2016aa_FL,Chen20FL}.
Essentially, the FEEL framework is a distributed implementation of {\it stochastic gradient decent} (SGD)  over wireless networks. A typical training process involves iterations between 1)  broadcasting of the global model under training from edge server to devices for local SGD execution using local data, and 2) local models/gradients uploading from devices to edge server for aggregation and global model updating. Although the uploading of high-volume raw data is avoided, the updates aggregation process in FEEL may still suffer from a communication bottleneck due to the high-dimensionality of each updates and the multiple access by many devices over wireless links. To tackle this issue, one promising solution called {\it over-the-air} FEEL (Air-FEEL) has been proposed, which exploits the {\it over-the-air computation} (AirComp) for ``one-shot" aggregation via concurrent update transmission, such that communication and computation are integrated in a joint design by exploiting  the superposition property of a {\it multiple access channel} (MAC)  \cite{Survey_FEEl,nomo_function_Nazer,Gastpar08}.

The idea of AirComp was first proposed in \cite{nomo_function_Nazer} in the context of data aggregation in sensor networks, where it is surprisingly found that ``interference" can be harnessed by structured codes to help functional computation over a MAC. Inspired by the finding, it was shown in the subsequent work \cite{Gastpar08} that for Gaussian {\it independent and identically distributed} (i.i.d.) data sources, the uncoded transmission is optimal in terms of distortion minimization. Besides the information-theoretic studies, various practical issues faced by AirComp implementation were also considered in \cite{Abari15,Cao_PowerTWC,Cao_2020aa}. In particular, the synchronization issue in AirComp was addressed in \cite{Abari15} via an innovative idea of shared clock broadcasting from edge server to devices. The optimal power control policies for AirComp over fading channels were derived in \cite{Cao_PowerTWC} to minimize the average computation distortion, and the cooperative interference management framework for coordinating coexisting AirComp tasks over multi-cell networks was developed in \cite{Cao_2020aa}.

More recently, AirComp found its merits in the new context of FEEL, known as Air-FEEL, for communication-efficient update aggregation as demonstrated in a rich set of prior works \cite{GZhu2020TWC,Amiri2020TSP,KYang2020TWC,NZhang2020Ar,GZhu2020Ar,DLiu2020Ar}. Specifically, a broadband Air-FEEL solution was proposed in \cite{GZhu2020TWC}, where several communication-learning tradeoffs were derived to guide the design of device scheduling. Around the same time, a source-coding algorithm exploiting gradient sparsification was proposed in \cite{Amiri2020TSP} to implement Air-FEEL with compressed updates for higher communication efficiency. In parallel, a joint design of device scheduling and beamforming in a multi-antenna system was presented in \cite{KYang2020TWC} to accelerate Air-FEEL. Subsequently, the gradient statistics aware power control was investigated in \cite{NZhang2020Ar} to further enhance the performance of Air-FEEL. Furthermore, to allow Air-FEEL compatible with digital chips embedded in modern edge devices, Air-FEEL based on digital modulation was proposed in \cite{GZhu2020Ar} featuring one-bit quantization and modulation at the edge devices and majority-vote based decoding at the edge server. Besides the benefit of low latency, Air-FEEL was also found to be beneficial in data privacy enhancement as individual updates are not accessible by edge server, eliminating the risk of model inversion attack  \cite{DLiu2020Ar}.

Despite the promise in high communication efficiency, Air-FEEL may suffer from severe learning performance degradation due to the aggregation error caused by the non-uniform channel fading over devices and noise perturbation. Prior work in this field mostly assumed channel inversion power control (or its variants) \cite{GZhu2020TWC,Amiri2020TSP,KYang2020TWC,DLiu2020Ar} in an effort to reducing the aggregation error by aligning the channel gains, which could be highly suboptimal in deep fading scenarios due to the noise amplification.  Although there exists one relevant study on power control for Air-FEEL system in \cite{NZhang2020Ar}, it focused on the minimization of the intermediate aggregation distortion (e.g., mean squared error) instead of the ultimate learning performance (e.g., the general loss function). Therefore, there still leaves a research gap in learning performance optimization of Air-FEEL by judicious power control, motivating the current work. To close the gap, we first analyze the convergence behavior of the Air-FEEL by deriving the optimality gap of the loss-function under arbitrary power control policy. Then the power control problem is formulated to minimize the optimality gap for convergence acceleration, subject to a set of average and maximum power constraints at edge devices. The problem is generally non-convex and challenging to solve due to the coupling of power control variables over different devices and iterations. The challenge is tackled by the joint use of {\it successive convex approximation} (SCA) and trust region methods in the optimized power control algorithm derivation. Numerical results show that the optimized power control policy achieves significantly faster convergence than the benchmark policies such as channel inversion and uniform power transmission, thus opening up a new degree-of-freedom for regulating the performance of Air-FEEL by power control.

\section{System Model}\label{sec:system}
\begin{figure}
\centering
 \setlength{\abovecaptionskip}{-4mm}
\setlength{\belowcaptionskip}{-4mm}
    \includegraphics[width=3.5in]{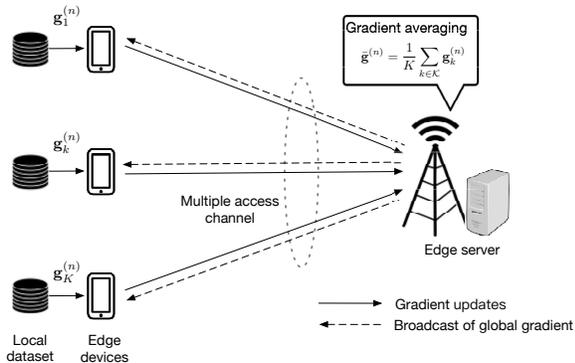}
\caption{Illustration of over-the-air federated edge learning. } \label{fig:model}
\end{figure}

We consider an Air-FEEL system consisting of an edge server and $K\ge 0$ edge devices, as shown in Fig.~\ref{fig:model}. With the coordination of the edge server, the edge devices cooperatively train a shared machine learning model via over-the-air update aggregation as elaborated in the sequel.

\subsection{Learning Model}
We assume that the learning model is represented by the parameter vector ${\bf w}\in\mathbb{R}^q$ with $q$ denoting the model size.
Let ${\mathcal D}_k$ denote the local dataset at edge device $k$, in which the $i$-th sample and its ground-true label are denoted by ${\bf x}_i$ and $y_i$, respectively.
Then the local loss function of the model vector $\bf w$ on ${\mathcal D}_k$ is
\begin{align}\label{LocalLossFunction}
F_k({\bf w})=\frac{1}{|{\mathcal D}_k|} \sum \limits_{({\bf x}_i,y_i)\in{\mathcal D}_k} f({\bf w},{\bf x}_i,y_i)+\rho R({\bf w}),
\end{align}
where $f({\bf w},{\bf x}_i,y_i)$ denotes the sample-wise loss function quantifying the prediction error of the model $\bf w$ on the sample ${\bf x}_i$ {\it with respect to} (w.r.t.) its ground-true label $y_i$, and $R({\bf w})$ denotes the strongly convex regularization function scaled by a hyperparameter $\rho\geq 0$.
For notational convenience, we simplify $f({\bf w},{\bf x}_i,y_i)$ as $f_i({\bf w})$.
Then, the global loss function on all the distributed datasets is given by
\begin{align}\label{GlobalLossFunction}
F({\bf w})=\frac{1}{K}\sum\limits_{k\in\mathcal K} D_k F_k({\bf w}),
\end{align}
where ${\mathcal D}=\cup_{k\in\mathcal K} {\mathcal D}_k$ with $D_{\rm tot}=|{\mathcal D} |$, and the sizes of datasets in all edge devices are assumed to be uniform for notation simplicity, i.e., $|{\mathcal D}_k|=\bar D,\forall k\in\mathcal K$.

The objective of the training process is to minimize the global loss function $F({\bf w})$:
 \begin{align}\label{OptimalParameter}
 {\bf w}^{\star}=\arg \min_ {\bf w} F({\bf w}).
\end{align}
Instead of directly uploading all the local data to the edge server for centralized training, the learning process in \eqref{OptimalParameter} can be implemented iteratively in a distributed manner based on gradient-averaging approach as illustrated in Fig.~\ref{fig:model}.

At each communication round $n$, the machine learning model is denoted by ${\bf w} ^{(n)}$. Then each edge device can compute the local gradient denoted by ${\bf g}_{k}^{(n)}$ using the local dataset ${\mathcal D}_k$:
\begin{align}\label{sys_LocalGradient}
{\bf g}_{k}^{(n)}=	\frac{1}{|{\mathcal D}_k|} \sum \limits_{({\bf x}_i,y_i)\in{\mathcal D}_k} \nabla f_i({\bf w}^{(n)})+\rho \nabla R({\bf w}),
\end{align}
where $\nabla$ is the gradient operator and we assume that the whole local dataset is used to estimate the local gradients.
Next, the edge devices upload all local gradients to the edge server, which are further averaged to obtain the global gradient:
\begin{align}\label{sys_GlobalGradient}
\bar{\bf g}^{(n)}=	\frac{1}{K}\sum\limits_{k\in\mathcal K} {\bf g}_{k}^{(n)}.
\end{align}
Then, the global gradient estimate is broadcast from edge server to edge devices, based on which edge device can update its own model under training via
\begin{align}\label{sys_ModelUpdate}
{\bf w}^{(n+1)}={\bf w}^{(n)}-\eta\cdot \bar{\bf g}^{(n)},
\end{align}
where $\eta$ is the learning rate. Notice that the above procedure continues  until convergence criteria is met or the maximum number of iterations is achieved.

 \subsection{Basic Assumptions on Learning Model}
 To facilitate the convergence analysis next, we make several standard assumptions on the loss function and gradient estimates.

 \begin{assumption}[Smoothness]\label{Assump_Smooth}\emph{
Let ${\bf g}=\nabla F({\bf w})$ denote the gradient of the loss function evaluated at point ${\bf w}$. Then there exists a non-negative constant vector  ${\bf L}\in\mathbb{R}^q$, such that
\begin{align*}
F({\bf w})\!-\!\left[ F({\bf w}^{\prime})\! +\! {\bf g}^T (\!{\bf w}\!\!-\! {\bf w}^{\prime})\right] \le \frac{1}{2}\sum_{i=1}^{q}\! L_i({{ w}_i\!-\!{w}^{\prime}_i}\!)^2, \forall {\bf w}, {\bf w}^{\prime},
\end{align*}
where the superscript $T$ denotes the transpose operation.}
\end{assumption}

\begin{assumption}[Polyak-Lojasiewicz Inequality]\label{Assump_PL}\emph{
Let $F^{\star}$ denote the optimal loss function value to problem \eqref{OptimalParameter}. There exists a constant $\mu\ge 0$ such that the global loss function $F({\bf w})$ satisfies the following Polyak-Lojasiewicz (PL) condition:
\begin{align*}
	\| {\bf g}\|_2^2 \ge 2\mu(F({\bf w})-F^{\star}).
\end{align*}
	}
\end{assumption}
 Notice that the above assumption is more general than the standard assumption of strong convexity \cite{Karimi2016}. Typical loss functions that satisfy the above two assumptions include logistic regression, linear regression and least squares.

 \begin{assumption}[Variance Bound]\emph{
The local gradient estimates $\{{\bf g}_k\}$, defined in \eqref{sys_LocalGradient}, where the index $(n)$ is omitted for simplicity, are assumed to be independent and unbiased estimates of the batch gradient ${\bf g}$ with coordinate bounded variance, i.e.,
\begin{align}
	&\mathbb{E}[{\bf g}_k]={\bf g}, \forall k\in\mathcal K,\\
	&\mathbb{E}[ ({g}_{k,i}-g_i)^2]\le \sigma_i^2, \forall k\in\mathcal K, \forall i,
\end{align}
where ${g}_{k,i}$ and $g_i$ are defined as the $i$-th element of $\{{\bf g}_k\}$ and ${\bf g}$, respectively, and ${\bm\sigma}=[\sigma_1,\cdots,\sigma_q]$ is a vector of non-negative constants.
 }
\end{assumption}

\vspace{-0.45cm}
\subsection{Communication Model}
\vspace{-0.2cm}

The distributed training latency is dominated by the update aggregation process, especially when the number of devices becomes large. Therefore, we focus on the aggregation process over a MAC. Instead of treating different devices' update as interference, we consider AirComp for fast update aggregation by exploiting the superposition property of MAC.
We assume that the channel coefficients remain unchanged within a communication round, and may change over different communication rounds.
Besides, the channel state information (CSI) is assumed to be available at all edge devices, so that they can perfectly compensate for the phases introduced by the wireless channels.

Let $\hat h_k^{(n)}$ denote the complex channel coefficient from device $k$ to the edge server at communication round $n$, and $h_k^{(n)}$ denote its magnitude with $h_k^{(n)}=|\hat h_k^{(n)}|$.
During the gradient-uploading phase, all the devices transmit simultaneously over the same time-frequency block, and thus the received aggregate signal is given by
 \begin{align}\label{sys_ReceivedSignal}
 	{\bf y}^{(n)}=\sum\limits_{k\in\mathcal K}h_k^{(n)}\sqrt{p_k^{(n)}}{\bf g}_{k}^{(n)}+{\bf z}^{(n)},
 \end{align}
 in which $p_k^{(n)}$ denotes the transmit power at device $k$, and ${\bf z}^{(n)}\in\mathbb{R}^q$ denotes the additive white Gaussian noise with ${\bf z}^{(n)}\sim{\mathcal CN}(0,N_0\bf I)$, where $N_0$ is the noise power density and $\bf I$ is an identity matrix. Therefore, the global gradient estimate at the edge server is given by
\begin{align}\label{sys_GlobalGradient}
	\hat{\bf g}^{(n)}=\frac{{\bf y}^{(n)}}{K}.
\end{align}
The devices can adaptively adjust their transmit powers for enhancing the learning performance.
In practice, the transmit power of each edge device $k\in\mathcal K$ at each communication round is constrained by a maximum power budget $\bar P_{k}$:
\begin{align}
p_{k}^{(n)}  \leq \bar{P}_{k},~\forall k\in{\mathcal K}, ~\forall n.\label{sys_bar_P_max}
\end{align}
In addition, each device $k\in\mathcal K$ is also constrained by an average power budget denoted by $\tilde{P}_{k}$ over the whole training period as expressed below:\begin{align}
\frac{1}{N}\sum \limits_{n\in\mathcal{N}}p_{k}^{(n)}  \leq \tilde{P}_{k},~\forall k\in{\mathcal K}.\label{sys_bar_P_ave}
\end{align}
Here, we generally have $\tilde{P}_{k} \le \bar{P}_{k},~\forall k\in{\mathcal K}$.

 \section{Convergence Analysis for Air-FEEL with Adaptive Power Control}

 In this section, we formally characterize the learning performance of Air-FEEL system, which is derived to be a function of transmit powers of all devices.


Let $N$ denote the number of needed communication rounds and $L\triangleq \|\bf L\|_{\infty}$. For notational convenience,  we use $F^{(n+1)}$ to represent $F({\bf w}^{(n+1)})$.
The optimality gap after $N$ communication rounds defined by $F^{(N+1)}-F^{\star}$ is derived in the following theorem, from which we can understand the convergence behavior of Air-FEEL.

\begin{theorem}[Optimality Gap]\label{ConvergenceRate}\emph{
The optimality gap for Air-FEEL, with arbitrary transmit power control policy $\{p_k^{(n)}\}$, is given as
\begin{align}
&\mathbb{E}\left[F^{(N+1)} \right]-F^{\star}\leq {\Phi}(\{p_k^{(n)}\},\eta)\notag\\
&\!\!\triangleq\!\prod_{n=1}^{N}\!A^{(n)}\!\!\left(F^{(1)}\!-\!F^{\star}\!\right) \!+\!\sum_{n=1}^{N-1}\!\left(\!\prod_{i=n+1}^{N}\!A^{(i)}\!\right)\!B^{(n)}\!+\!B^{(N)},\label{Conv_F_Gap}\!
\end{align}
with $A^{(n)}=1-\frac{2\mu\eta}{K}\sum\limits_{k\in\mathcal K}\left(h_k^{(n)}\sqrt{p_k^{(n)}}-\frac{ \eta L}{2K}(h_k^{(n)})^2p_k^{(n)}\right) $ and $B^{(n)}=\frac{ \eta^2L\|{\bm \sigma}\|_2^2}{2K^2}\left(\sum\limits_{k\in\mathcal K}(h_k^{(n)})^2p_k^{(n)}\right)+\frac{ \eta^2LN_0q}{2K^2}$.
	}
\end{theorem}
\begin{IEEEproof}
The proof follows the widely-adopted strategy of relating the norm of the gradient to the expected improvement made in a single algorithmic step, and comparing this with the total possible improvement.
\begin{footnotesize}
\begin{align*}
	&F^{(n+1)}- F^{(n)}\\
&\overset{(a)}{\leq}   ({\bf g}^{(n)})^T ({\bf w}^{(n+1)}- {\bf w}^{(n)}) + \frac{1}{2}\sum_{i=1}^{q} L_i({{ w}_i^{(n+1)}-{w}^{(n)}_i})^2,\\
	&\overset{(b)}{\leq} ({\bf g}^{(n)})^T ({\bf w}^{(n)}-\eta\cdot \hat{\bf g}^{(n)}- {\bf w}^{(n)}) + \frac{L}{2}\|{\bf w}^{(n)}-\eta\cdot \hat{\bf g}^{(n)}- {\bf w}^{(n)}\|_2^2\\
	&=-\eta ({\bf g}^{(n)})^T \hat{\bf g}^{(n)}+ \eta^2\frac{L}{2}\|\hat{\bf g}^{(n)}\|_2^2\\
	&\!\!=\!\!-\!\frac{\eta}{K} ({\bf g}^{(\!n\!)}\!)^T\!\!\left(\!\sum\limits_{k\in\mathcal K}\!h_k^{(\!n\!)}\sqrt{p_k^{(\!n\!)}}\!{\bf g}_{k}^{(\!n\!)}\!\!+\!\!{\bf z}^{(\!n\!)}\!\right)\!\!\!+\!\frac{ \eta^2L}{2K^2}\!\!\left\|\!\sum\limits_{k\in\mathcal K}\!\!h_k^{(\!n\!)}\sqrt{p_k^{(\!n\!)}}{\bf g}_{k}^{(\!n\!)}\!+\!{\bf z}^{(\!n\!)}\!\right\|_2^2\!,\!
\end{align*}
\end{footnotesize}
where the inequalities (a) and (b) follows the Assumption~\ref{Assump_Smooth} and $L\triangleq \|\bf L\|_{\infty}$.
By subtracting $F^{\star}$ and taking expectation at both sides, the convergence rate of each communication round is given by \eqref{App_Con_expectation}.
Next, \eqref{App_Con_gap} is obtained by applying the PL condition in the Assumption~\ref{Assump_PL}.
Then, by applying above inequality repeatedly through $N$ iterations, after some simple algebraic manipulation we have \eqref{Conv_F_Gap}, which completes the proof.

\begin{figure*}[t]
\begin{align}
&\mathbb{E}\left[F^{(n+1)} \right]-F^{\star}\notag\\
	&\le F^{(n)}-F^{\star}-\frac{\eta}{K}\left(\sum\limits_{k\in\mathcal K}h_k^{(n)}\sqrt{p_k^{(n)}}\right)\|{\bf g}^{(n)}\|_2^2 +\frac{ \eta^2L}{2K^2}\left(\sum\limits_{k\in\mathcal K}(h_k^{(n)})^2p_k^{(n)}\right)\left(\left\|{\bf g}^{(n)}\right\|_2^2+\left\|{\bm \sigma}\right\|_2^2\right)+\frac{ \eta^2LN_0q}{2K^2}\notag\\
		&= F^{(n)}-F^{\star}-\left[\sum\limits_{k\in\mathcal K}\left(\frac{\eta}{K}h_k^{(n)}\sqrt{p_k^{(n)}}-\frac{ \eta^2L}{2K^2}(h_k^{(n)})^2p_k^{(n)}\right)\right]\|{\bf g}^{(n)}\|_2^2 +\frac{ \eta^2L}{2K^2}\left(\sum\limits_{k\in\mathcal K}(h_k^{(n)})^2p_k^{(n)}\right)\left\|{\bm \sigma}\right\|_2^2+\frac{ \eta^2LN_0q}{2K^2}.\label{App_Con_expectation}
\end{align}
\vspace*{-0.2\baselineskip}
\end{figure*}

\begin{figure*}
\begin{align}
\!\!\mathbb{E}\!\left[\!F^{(n+\!1)} \!\right]\!\!-\!F^{\star}\!\!\le\! \! \underbrace{ \left[\!1\!-\!2\mu \! \! \left(\! \sum\limits_{k\in\mathcal K}\!\!\left(\!\frac{\eta}{K}h_k^{(n)}\sqrt{p_k^{(n)}}\!-\!\frac{ \eta^2L}{2K^2}(h_k^{(n)})^2p_k^{(n)}\right)\! \! \right)\!  \right]\! }_{A^{(n)}}\left(\!F^{(n)}-F^{\star}\!\right) 	\! +\underbrace{\frac{ \eta^2L\left\|{\bm \sigma}\right\|_2^2}{2K^2}\sum\limits_{k\in\mathcal K}(h_k^{(n)})^2p_k^{(n)}\!+\!\frac{ \eta^2LN_0q}{2K^2}}_{B^{(n)}}.\label{App_Con_gap}
\end{align}\!
\vspace*{-1\baselineskip}
\end{figure*}
\end{IEEEproof}



Further applying the mean inequality $(a_1a_2\cdots a_m)\leq(\frac{a_1+a_2+\cdots +a_m}{m})^m$, we can derive a more elegant upper bound for the expression in \eqref{Conv_F_Gap} to attain more insights as follows
\begin{align}
& {\Phi}(\{p_k^{(n)}\},\eta)\leq\!\!\alpha^N\left(\!F^{(\!1\!)}\!-\!F^{\star}\!\right) \!\!+\!\!\sum_{n=1}^{N}\!\!B^{(n)}\beta_{(n)}^{N-n},\label{Conv_F_Gap_bound}
\end{align}
where $\alpha=\frac{\sum_{i=1}^{N}\!\!A^{(i)}}{N}$ and $\beta_{(n)}=\frac{\sum_{i=n+1}^{N}A^{(i)}}{N-n}$ for $n=1,\cdots,N-1$ while $\beta_{(N)}=1$.
\begin{remark}\emph{
	The first term on the right hand side of \eqref{Conv_F_Gap_bound} suggests that the effect of initial optimality gap vanishes as the number of communication round $N$ increases. The second term reflects the impact of the power control and additive noise power on the convergence process, that is, transmission with more power in the initial learning iterations is more beneficial in decreasing the optimality gap. This is because that the contribution of power control at iteration $n$ is discounted by a factor $\beta_{(n)}^{N-n}$.}
\end{remark}

 \section{Power Control Optimization}
 In this section, we focus on speeding up the convergence rate by minimizing the optimality gap in Theorem~\ref{ConvergenceRate}, under the power constraints stated in \eqref{sys_bar_P_max} and \eqref{sys_bar_P_ave}.
The optimization problem is thus formulated as
 \begin{align}
 	\mathbf{P1:} \min_{\{p_k^{(n)}\ge 0\},\eta\ge 0} ~~&{\Phi}(\{p_k^{(n)},\eta\})\notag\\
 {\rm s.t.}~~~~~&\eqref{sys_bar_P_max}~\text{and}~\eqref{sys_bar_P_ave}.\notag
 \end{align}
Due to the coupling between the power control $\{p_k^{(n)}\}$ and learning rate $\eta$, problem (P1) is non-convex and hard to solve. We resort to the alternating optimization technique for efficiently solving this problem. In particular, we first solve problem (P1) under any given $\eta$, and then apply a one-dimension search to find the optimal $\eta$ that achieves the minimum objective value.

Let $\tilde{\Phi}(\{p_k^{(n)}\})={\Phi}(\{p_k^{(n)},\eta\})$ under any given $\eta$.
 Note that the transmit powers at different devices and different communication rounds are coupled with each other in the objective function in \eqref{Conv_F_Gap} under given learning rate $\eta$, leading to a highly non-convex problem:
  \begin{align}
 	\mathbf{P2:} \min_{\{p_k^{(n)}\ge 0\}} ~~& \tilde{\Phi}(\{p_k^{(n)}\})\notag\\
 {\rm s.t.}~~~~~&\eqref{sys_bar_P_max}~\text{and}~\eqref{sys_bar_P_ave}.\notag
 \end{align}
To tackle this problem, we propose an iterative algorithm to obtain an efficient solution using the SCA technique. The key idea is that under any given local point at each iteration, we can approximate the non-convex objective as a constructed convex one. Therefore, after solving a series of approximate convex problems iteratively, we can obtain a high-quality suboptimal solution to problem (P2).

Let $\{p_k^{(n)}[i]\}$ denote the local point at the $i$-th iteration with $i\ge 0$, and ${\mathcal N}\triangleq \{1,\cdots,N\}$ the set of communication rounds. Notice that by checking the first-order Taylor expansion of $\tilde{\Phi}(\{p_k^{(n)}\})$ w.r.t. $\{p_k^{(n)}\}$ at the local point $\{p_k^{(n)}[i]\}$, it follows that
\begin{align*}
	&\tilde{\Phi}(\{p_k^{(n)}\})\approx \bar{\Phi}(\{p_k^{(n)}\})\notag\\
	&~~ \triangleq \!\tilde{\Phi}(\{p_k^{(n)}[i]\})\!+\!\sum\limits_{n\in{\mathcal N}} \!\sum\limits_{k\in{\mathcal K}}\!\left(p_k^{(n)}-p_k^{(n)}[i]\right)\nabla \tilde{\Phi}(\{p_k^{(n)}[i]\}),
\end{align*}
where $\nabla \tilde{\Phi}(\{p_k^{(n)}[i]\})$ represents the first-order derivative w.r.t. $p_k^{(n)}[i]$, given in \eqref{Derivation_1} and \eqref{Derivation_2}.
\begin{figure*}[t]
\begin{align}
	\nabla \tilde{\Phi}(p_k^{(n)}[n])&=-\frac{\mu\eta h_k^{(n)}\left(F^{(1)}-F^{\star}\right)}{K}\left(\frac{1}{\sqrt{p_k^{(n)}}}-\frac{\eta L h_k^{(n)}}{K}\right)\prod_{i\in{\mathcal N}\setminus \{n\}}A^{(i)}+\frac{\eta^2L \left\|{\bm \sigma}\right\|_2^2 (h_k^{(n)})^2\prod_{j=n}^{N}A^{(j)}}{2K^2A^{(n)}}\notag\\
	&~~-\frac{\mu\eta h_k^{(n)}}{K}\left(\frac{1}{\sqrt{p_k^{(n)}}}-\frac{\eta L h_k^{(n)}}{K}\right)\sum_{\ell=1}^{n-1} B_{(\ell)}\frac{\prod_{j=\ell}^{N}A^{(j)}}{A^{(n)}A^{(\ell)}},\forall n\in\mathcal{N}\setminus \{1\}\label{Derivation_1}\\
	\nabla \tilde{\Phi}(p_k^{(1)}[i])&=-\frac{\mu\eta h_k^{(1)}\left(F^{(1)}-F^{\star}\right)}{K}\left(\frac{1}{\sqrt{p_k^{(1)}}}-\frac{\eta L h_k^{(1)}}{K}\right)\prod_{i\in{\mathcal N}\setminus \{1\}}A^{(i)}+\frac{\eta^2L \left\|{\bm \sigma}\right\|_2^2 (h_k^{(1)})^2}{2K^2}\prod_{i\in{\mathcal N}\setminus \{1\}}A^{(i)}.\label{Derivation_2}
\end{align}
\vspace*{-1\baselineskip}
\end{figure*}\!

In this case, $\bar{\Phi}(\{p_k^{(n)}\})$ is linear w.r.t. $\{p_k^{(n)}\}$.
To ensure the approximation accuracy, a series of trust region constraints are imposed as \cite{YLiu_Trust}
\begin{align}
	|p_{k}^{(n)}[i]-p_{k}^{(n)}[i-1]|\le \Gamma[i], ~\forall k\in\mathcal{K}, \forall n\in\mathcal{N},\label{TrustRegion}
\end{align}
where $\Gamma[i]$ denotes the radius of the trust region.
By replacing $\bar{\Phi}(\{p_k^{(n)}\})$ as the approximation of $\tilde{\Phi}(\{p_k^{(n)}\})$ and introducing an auxiliary variable $\gamma$, the approximated problem at the $i$-th iteration is derived as a convex problem:
 \begin{align}
 	\mathbf{P2.1:} \min_{\{p_k^{(n)}[i]\},\gamma\ge 0} ~~&\gamma \notag\\
{\rm s.t.}~~~~~&\bar{\Phi}(\{p_k^{(n)}[i]\})\le \gamma\\
&\eqref{sys_bar_P_max},~\eqref{sys_bar_P_ave},~\text{and}~\eqref{TrustRegion},\notag
 \end{align}
which can be directly solved by CVX \cite{cvx}.

Let $\{p_k^{(n)*}[i]\}$ denote the optimal power control policy to problem (P2.1) at local point $\{p_k^{(n)}[i]\}$. Then, we can obtain an efficient iterative algorithm to solve problem (P2) as follows. In each iteration $i \ge 1$, the power control is updated as $\{p_k^{(n)*}[i]\}$ by solving problem (P2.1) at local point $\{p_k^{(n)}[i]\}$, i.e. $p_k^{(n)}[i+1]=p_k^{(n)*}[i],\forall n\in\mathcal N, \forall k\in\cal K$, where $\{p_k^{(n)}[0]\}$ denotes the initial power control.
At the $i$-th iteration, we compute the objective value in problem (P2) by replacing $\{\hat{p}_k^{(n)*}[i]\}$ as $\{p_k^{(n)*}\}$. If the objective value decreases, we then replace the current point by the obtained solution and go to the next iteration; otherwise, we update $\Gamma[i]=\Gamma[i]/2$ and continue to solve problem (P2.1). This algorithm would stop until that $\Gamma[i]$ is lower than a given threshold denoted by $\epsilon$.
In summary, the proposed algorithm is presented in Algorithm 1.

\begin{table}[htp]
\begin{center}\vspace{-0.1cm}
\hrule
\vspace{0.2cm} \textbf{Algorithm 1 for Solving Problem (P2)}\vspace{0.2cm}
\hrule \vspace{0.1cm} 
\begin{itemize}
    \item[1]  Initialization: Given the initial power control $\{p_k^{(n)}[0]\}$; let $i=0$.
    \item[2]  {\bf Repeat:}
                \begin{itemize}
                \item[a)]  Solve problem (P1.1) under given $\{p_k^{(n)}[i]\}$ to obtain the optimal solution as $\{p_k^{(n)*}[i]\}$;
                 \item[b)]  If the objective value of problem (P2) $\tilde{\Phi}(\{p_k^{(n)}\})$ decreases, then update $p_k^{(n)}[i+1]=p_k^{(n)*}[i],\forall n\in\mathcal N$ with $i=i+1$; otherwise $\Gamma[i]=\Gamma[i]/2$;
                \end{itemize}
     \item[3] {\bf Until} $\Gamma[i]\le \epsilon$.
    \end{itemize}
\hrule \vspace{0cm}
\end{center}\vspace{-0.5cm}
\end{table}

With the obtained power control in Algorithm 1, we can find the optimal $\eta$ accordingly via a one-dimensional search.

   \vspace{-0.1cm}
\section{Simulation Results}\label{sec_simu}

In this section, we provide simulation results to validate the performance of the proposed power control policy for Air-FEEL.
In the simulation, the wireless channels from each device to the edge server over fading states follow i.i.d. Rayleigh fading, such that $h_k$'s are modeled as i.i.d. {\it circularly symmetric complex Gaussian} (CSCG) random variables with zero mean and unit variance.
The dataset with size $D_{\rm tot}=600$ at all device are randomly generated, where part of the data, namely $100$ pairs (${\bf x}$, $y$), are left for prediction, and the remaining ones are used for model training. The generated data sample vector ${\bf x}$ follow i.i.d. Gaussian distribution as $\mathcal{N}(0,{\bf I})$ and the label $y$ is obtained as $y=x(2)+3x(5)+0.2z$, where $x(t)$ represents the $t$-entry in vector ${\bf x}$ and $z$ is the observation noise with i.i.d. Gaussian distribution, i.e., $z\sim\mathcal{N}(0,1)$.
Unless stated otherwise, the data samples are evenly distributed among the $K=20$ devices, and thus it follows $D_k=25$.
Moreover, we apply ridge regression with the sample-wise loss function $f({\bf w},{\bf x},y)=\frac{1}{2}\| {\bf x}^T{\bf w}-y\|^2$ and the regularization function $R({\bf w})=\|{\bf w}\|^2$ with $\rho=5\times 10^{-5}$ in this paper.
Furthermore, recall that $D_{\rm tot}=\sum_{k\in \mathcal{K}}D_k$ and then we can obtain the smoothness parameter $L$ and PL parameter $\mu$ as the largest and smallest eigenvalues of the data Gramian matrix ${\bf X}^T{\bf X}/D_{\rm tot}+10^{-4}{\bf I}$, in which ${\bf X}=[{\bf x}_1,\cdots,{\bf x}_{D_{\rm tot}}]^T$ is the data matrix. The optimal loss function $F^{\star}$ is computed according to the optimal parameter vector $\bf w^{\star}$ to the learning problem \eqref{OptimalParameter}, where ${\bf w}^{\star}=({\bf X}^T{\bf X}+\rho{\bf I})^{-1}{\bf X}^T{\bf y}$ with ${\bf y}=[y_1,\cdots,y_{D_{\rm tot}}]^T$.
We set the initial parameter vector as an all-zero vector and the noise variance $N_0=0.1$.

We consider two benchmark schemes for performance comparison, namely the {\it uniform power transmission} that transmits with uniform power over different communication round under the constraint of average power budget, and the {\it channel inversion} adopted in \cite{DLiu2020Ar}.
As for the performance metric for comparison, we consider the optimality gap and prediction error to evaluate the learning performance. 


%

\begin{figure}[htbp]  \vspace{-0.05cm}
  \centering
  \subfigure[Optimality gap versus varying number of devices.]
  {\label{fig:FL_v_K1}\includegraphics[width=8cm]{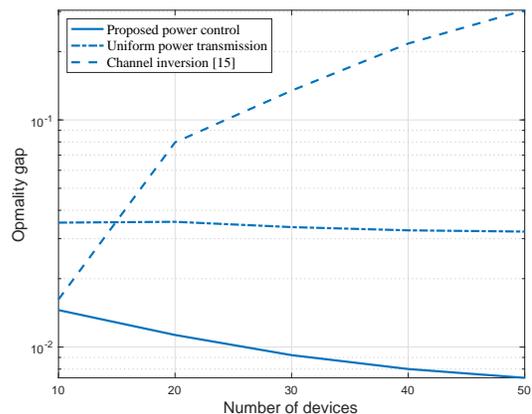}}
  \subfigure[Prediction error versus varying number of devices.]
  {\label{fig:FL_v_K2}
\includegraphics[width=8cm]{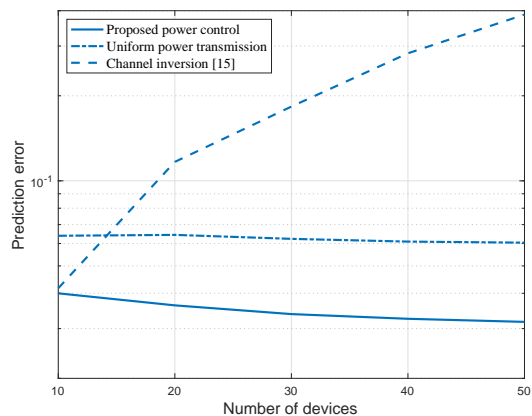}}
  \caption{Effect of number of devices on the learning performance of  Air-FEEL.}
  \label{Fig:FL_v_K}
\end{figure}

The effect of device population on learning performance is illustrated in Fig.~\ref{Fig:FL_v_K}  with $N=30$, where the power budgets at all devices are identically set to be  $\tilde P=1$ W and $\bar P=5$ W.
Notice that the increasing of device population may introduce both the  positive and negative effects on the learning performance. The positive effect is that the training process can exploit more data, while the negative effect is the increased aggregation error raised by AirComp over more devices. As observed in Fig.~\ref{Fig:FL_v_K}, the positive effect can be cancelled or even overweighed by the negative effect when applying the channel inversion or uniform power control. The blessing of including more devices in Air-FEEL can dominate the curse it brings only when the power control is judiciously optimized, showing the crucial role of power control in determining the the learning performance of Air-FEEL.

\begin{figure}[htbp] \vspace{-0.05cm}
  \centering
  \subfigure[Tendency of loss function.]
  {\label{fig:MSE_v_Con1}\includegraphics[width=8cm]{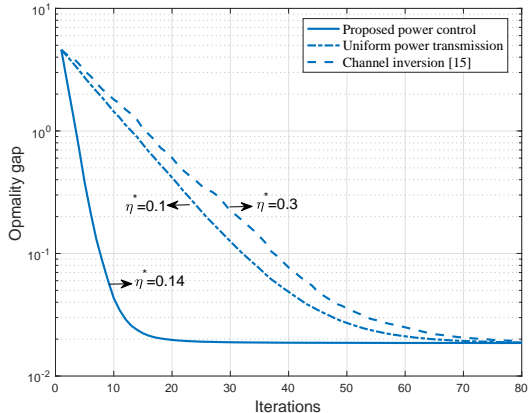}}
  \subfigure[Tendency of prediction error.]
  {\label{fig:MSE_v_Con2}
\includegraphics[width=8cm]{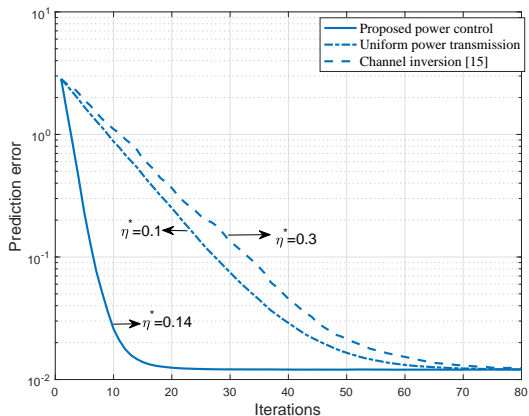}}
  \caption{Learning performance of Air-FEEL over iterations, where $\eta^*$ denotes the optimized learning rate after one-dimensional search.}
  \label{Fig:MSE_v_Con}
\end{figure}

Fig.~\ref{Fig:MSE_v_Con} shows the learning performance during the learning process under the optimized learning rate, where we set $K=20$, $\tilde P=1$ W, $\bar P=5$ W, and $N=80$.
It is observed that the proposed power control scheme can achieve faster convergence than both the channel-inversion and uniform-power-control schemes. This is attributed to the power control optimization directly targeting convergence acceleration.

\begin{figure}   \vspace{-0.1cm}
\centering
 \setlength{\abovecaptionskip}{-1mm}
\setlength{\belowcaptionskip}{-1mm}
    \includegraphics[width=3.5in]{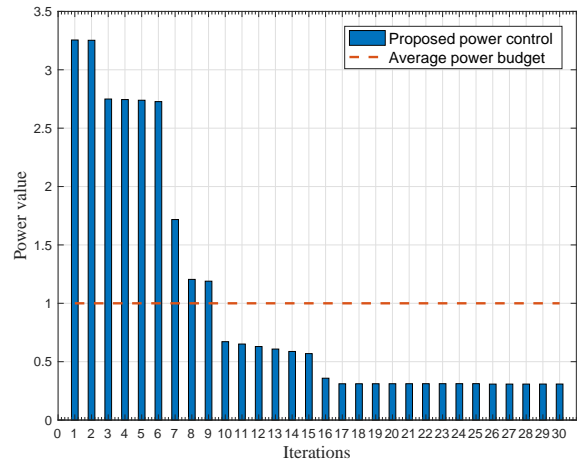}
\caption{The optimized power allocation over iterations under static channels.} \label{fig:MSE_V_PA}
\end{figure}

Fig.~\ref{fig:MSE_V_PA} shows the power allocation over a static channel with uniform channel gain during the learning process, where we set $K=20$, $\tilde P=1$ W, $\bar P=5$ W, and $N=30$.
It is observed that the power allocation over a static channel follows a stair-wise monotonously decreasing function.
The behavior of power control coincides the analysis on Remark 1.

\section{Conclusion}\vspace{-0.2cm}
In this paper, we exploit power control as a new degree of freedom to optimize the learning performance of Air-FEEL, a promising communication-efficient solution towards edge intelligence. To this end, we first analyzed the convergence rate of the Air-FEEL by deriving the optimality gap of the loss-function under arbitrary power control policy. Then the formulated power control problem aimed to minimize the optimality gap for accelerating convergence, subject to a set of average and maximum power constraints at edge devices. Due to the coupling of power control variables over different devices and iterations, the challenge of the formulated power control problem was tackled by the joint use of SCA and trust region methods. Numerical results demonstrated that the optimized power control policy can achieve significantly faster convergence than the benchmark policies such as channel inversion and uniform power transmission.

\vspace{-0.1cm}

\bibliography{AirCompforFL}
\bibliographystyle{IEEEtran}

\end{document}